\documentclass[journal,12pt,onecolumn]{IEEEtran}

\usepackage{amsfonts,amssymb,amsmath,epsfig}

\newcommand{\ba}{\left( \begin{array}}
\newcommand{\ea}{\end{array} \right)}
\newcommand{\bq}{\begin{eqnarray*}}
\newcommand{\eq}{\end{eqnarray*}}
\newcommand{\bqn}{\begin{eqnarray}}
\newcommand{\eqn}{\end{eqnarray}}

\usepackage{hyperref}
\hypersetup{
    colorlinks=true,%
    citecolor=blue,%
    filecolor=blue,%
    linkcolor=red,%
    urlcolor=blue}

\begin{document}

\title{Heat Kernel Smoothing in Irregular Image Domains}
\author{ Moo K. Chung$^1$, Yanli Wang$^2$,  Gurong Wu$^3$\\
\vspace{0.3cm}
$^1$University of Wisconsin, Madison, USA\\
Email: \url{mkchung@wisc.edu}\\
\vspace{0.2cm}
$^2$Institute of Applied Physics and Computational Mathematics, Beijing, China\\
Email: \url{wangyanliwyl@gmail.com}\\
\vspace{0.2cm}
       $^3$University of North Carolina, Chapel Hill, USA\\
       Email: \url{guorong_wu@med.unc.edu}
       }

\maketitle

\begin{abstract}
We present the discrete version of heat kernel smoothing on graph data structure. 
The method is used to smooth data in an irregularly shaped domains in 3D images. 
 New statistical properties are derived. As an application, we show how to filter out data in the lung blood vessel trees obtained from computed tomography. The method can be further used in representing the complex vessel trees
parametrically and extracting the skeleton representation of the trees.
\end{abstract}

\section{Introduction}
{\em Heat kernel smoothing} was originally introduced in the context of filtering out surface data defined on mesh vertices obtained from 3D medical images \cite{chung.2005.IPMI,chung.2005.NI}. The formulation uses the tangent space projection in approximating the heat kernel by iteratively applying Gaussian kernel with smaller bandwidth. Recently proposed spectral formulation to heat kernel smoothing \cite{chung.2015.MIA} constructs the heat kernel analytically using the eigenfunctions of the Laplace-Beltrami (LB) operator, avoiding the need for the linear approximation used in \cite{chung.2005.IPMI,han.2006}. 
	
In this paper, we present the discrete version of heat kernel smoothing on graphs.  Instead of Laplace-Beltrami operator, graph  Laplacian is used to construct the discrete version of heat kernel smoothing. New statistical properties are derived for kernel smoothing that utilizes the fact heat kernel is a probability distribution. Heat kernel smoothing is used to smooth out data defined on irregularly shaped domains in 3D images. 

The connection between the eigenfunctions of continuous and discrete Laplacians has been well established by several studies \cite{gladwell.2002,tlusty.2007}. Although many have introduced the discrete version of heat kernel in computer vision and machine learning, they mainly used the heat kernels to compute shape descriptors  or to define a multi-scale metric \cite{belkin.2006,sun.2009,bronstein.2010, deGoes.2008}. These studies did not use the heat kernels in filtering out  data on graphs. There have been significant developments in kernel methods in the machine learning community \cite{bernhard.2002,nilsson.2007,shawe.2004,steinke.2008,yger.2011}. However, to the best of our knowledge, the heat kernel has never been used in such frameworks. Most kernel methods in machine learning deal with the linear combination of kernels as a solution to penalized regressions. On the other hand, our kernel method does not have a penalized cost function. 
	
As a demonstration, the method is applied in irregularly shaped lung blood vessel trees obtained from computed tomography (CT). 3D image volumes are represented as a large 3D graph by connecting neighboring voxels in the vessel trees. Since heat kernel smoothing is analytically represented, it is possible to use the technique resample lung vessel trees in a higher resolution and obtain the skeleton representation of the trees for further shape analysis \cite{lindvere.2013,cheng.2014}.

\section{Preliminary}
Let $G = \{V,E\}$ be a graph with vertex set $V$ and edge set $E$. We will simply index the node set as $V = \{1,2 ,\cdots, p\}$. If two nodes $i$ and $j$ form an edge, we denote it as $i \sim j$. Let $W=(w_{ij})$ be the edge wight. The adjacency matrix of $G$ is often used as the edge weight. Various forms of graph Laplacian have been proposed \cite{chung.1997} but the most often used standard form $L=(l_{ij})$ is given by 
$$l_{ij} =\left(
\begin{array}{cc}
  -w_{ij}, &  i \sim j\\
  \sum_{i \neq j} w_{ij}, &  i = j \\
  0, & \mbox{ otherwise}      
\end{array}\right.$$
The graph Laplacian $L$ can then be written as $$L = D- W,$$ where $D=(d_{ij})$ is the diagonal matrix with $d_{ii} = \sum_{j =1}^n w_{ij}$. 
For this study, we will simply use the adjacency matrix so that the edge weights $w_{ij}$ are either 0 or 1. 

Unlike the continuous Laplace-Beltrami operators that may have possibly infinite number of eigenfunctions, we have up to $p$ number of eigenvectors $\psi_1, \psi_2, \cdots, \psi_p$ satisfying 
\bqn L \psi_j = \lambda_j \psi_j \label{eq:graph-eigen} \eqn
with 
$$0 = \lambda_1 < \lambda_2 \leq \cdots \leq \lambda_p.$$
The eigenvectors are orthonormal, i.e.,
$\psi_i^\mathsf{T} \psi_j  = \delta_{ij},$
the Kroneker's delta. The first eigenvector is trivially given as
$\psi_1 = {\bf 1}/\sqrt{p}$ with ${\bf 1} = (1, 1, \cdots, 1)^\mathsf{T}.$ 

All other higher order eigenvalues and eigenvectors are unknown analytically and have to be computed numerically. Using the eigenvalues and eigenvectors, the graph Laplacian can be decomposed spectrally. From (\ref{eq:graph-eigen}), 
\bqn L \Psi = \Psi \Lambda, \label{eq:graph-psi}\eqn
where  $\Psi = [\psi_1, \cdots, \psi_p]$ and $\Lambda$ is the diagonal matrix with entries  $\lambda_1, \cdots, \lambda_p$. 
Since $\Psi$ is an orthogonal matrix, 
$$\Psi \Psi^\mathsf{T} = \Psi^\mathsf{T}\Psi = \sum_{j=1}^p \psi_j \psi_j^\mathsf{T} = I_p,$$ the identify matrix of size $p$. Then (\ref{eq:graph-psi}) is written as
$$L  = \Psi \Lambda \Psi^\mathsf{T} = \sum_{j=1}^p \lambda_j \psi_j \psi_j^\mathsf{T}.$$
This is the restatement of the singular value decomposition (SVD) for Laplacian. 

For measurement  vector $f = (f_1, \cdots, f_p)'$ observed at the $p$ nodes, the discrete Fourier series expansion is given by
$$f = \sum_{j=1}^n \tilde f_j \psi_j,$$
where $\tilde f_j =  f^\mathsf{T}\psi_j = \psi_j^\mathsf{T}f$ are Fourier coefficients.

\begin{figure}[t]
\centering
\includegraphics[width=1\linewidth]{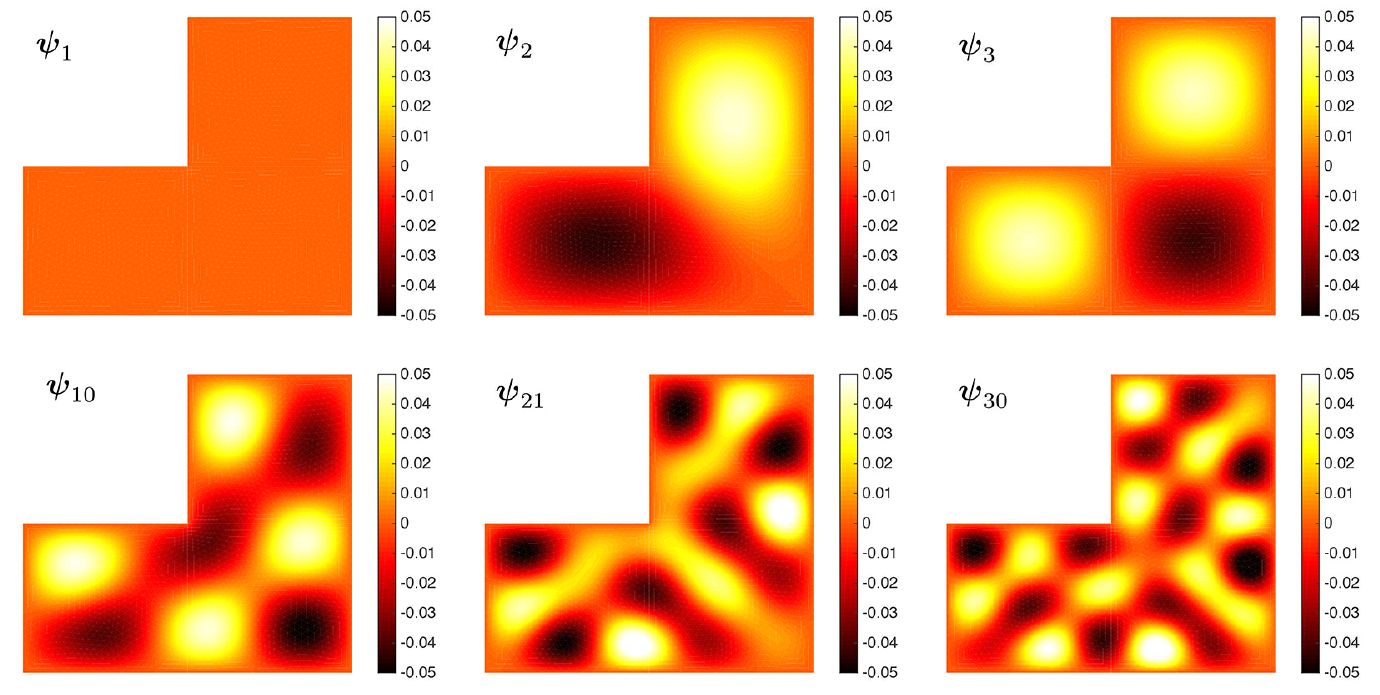}
\caption{First few eigenvector of the Laplacian in a L-shaped domain. The first eigenvector $\psi_0=0$}
\label{fig:runtime}
\end{figure}

\section{Heat kernel on graphs}
The {\em discrete heat kernel} $K_{\sigma}$ is a positive definite symmetric matrix of size $p \times p$ given by
\bqn K_{\sigma} = \sum_{j=1}^p e^{-\lambda_j \sigma} \psi_j\psi_j^\mathsf{T}, \label{eq:dhk} \eqn
where $\sigma$ is called the bandwidth of the kernel. Alternately, we can write (\ref{eq:dhk}) as
$$K_{\sigma} = \Psi e^{-{\sigma}\Lambda } \Psi^\mathsf{T},$$
where  $e^{-{\sigma}\Lambda }$ is the matrix logarithm of $\Lambda$. To see positive definiteness of the kernel,  for any nonzero $x \in \mathbb{R}^p$,
\bq
x^\mathsf{T} K_{\sigma} x &=& \sum_{j=1}^p e^{-\lambda_j \sigma} x^\mathsf{T}\psi_j\psi_j^\mathsf{T}x  \\
                &=& \sum_{j=1}^p e^{-\lambda_j \sigma} ( \psi_j^\mathsf{T}x )^2 > 0.\eq
When $\sigma=0$, $K_{0} = I_p$, identity matrix. When $\sigma=\infty$, by interchanging the sum and the limit, we obtain
$$K_{\infty} =  \psi_1\psi_1^\mathsf{T} = {\bf 1} {\bf 1}^\mathsf{T}/p.$$
$K_{\infty}$ is a degenerate case and the kernel is no longer positive definite. Other than these specific cases, the heat kernel is not analytically known in arbitrary graphs.

Heat kernel is doubly-stochastic \cite{chung.1997} so that 
$$K_{\sigma} {\bf 1} = {\bf 1}, \; {\bf 1}^\mathsf{T} K_{\sigma} = {\bf 1}^\mathsf{T}.$$
Thus, $K_{\sigma}$ is a probability distribution along columns or rows.

Just like the continuous counterpart,  the discrete heat kernel is also multiscale and has the scale-space property. Note
\bq K_{\sigma}^2 &=& \sum_{i,j=1}^p e^{-(\lambda_i + \lambda_j) \sigma} \psi_i\psi_i^\mathsf{T} \psi_j\psi_j^\mathsf{T}\\
&=& \sum_{j=1}^p e^{-2\lambda_j \sigma} \psi_j\psi_j^\mathsf{T} = K_{2\sigma}.
\eq
We used the orthonormality of eigenvectors. Subsequently, we have 
$$K_{\sigma}^n = K_{n\sigma}.$$

\begin{figure}[t]
\centering
\includegraphics[width=1\linewidth]{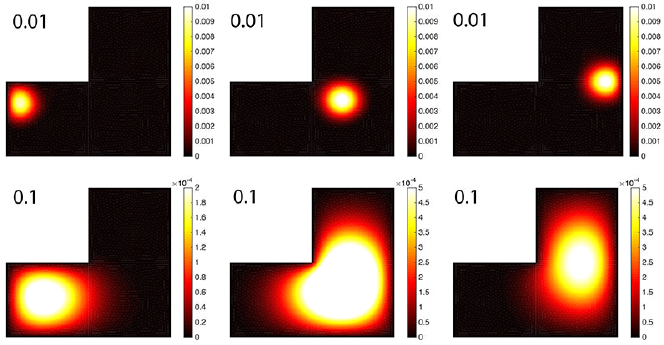}
 \caption{Heat kernels with bandwidth $\sigma =0.01, 0.1$. We have used degree 70 expansions but the shape is almost identical if we use higher degree expansions. The heat kernel is a probability distribution that follows the shape of domain.}
\label{fig:runtime}
 \end{figure}
 
\section{Heat kernel smoothing on graphs}
Discrete heat kernel smoothing of measurement vector $f$  is then defined as
\bqn K_{\sigma} * f = K_{\sigma}f  =\sum_{j=0}^p e^{-\lambda_j \sigma} {\tilde f}_j \psi_j, \label{eq:DHK} \eqn
This is the discrete analogue of heat kernel smoothing first defined in \cite{chung.2005.IPMI}. In discrete setting, the convolution $*$ is simply a matrix multiplication. Then $$K_{0} *f = f$$ and 
$$K_{\infty} *f =  \bar f {\bf 1},$$
where $\bar f = \sum_{j=1}^p f_j/p$ is the mean of signal $f$ over every nodes. When the bandwidth is zero, we are not smoothing data. As the bandwidth increases, the smoothed signal converges to the sample mean of all values.

\begin{figure*}[t]
\centering
\includegraphics[width=1\linewidth]{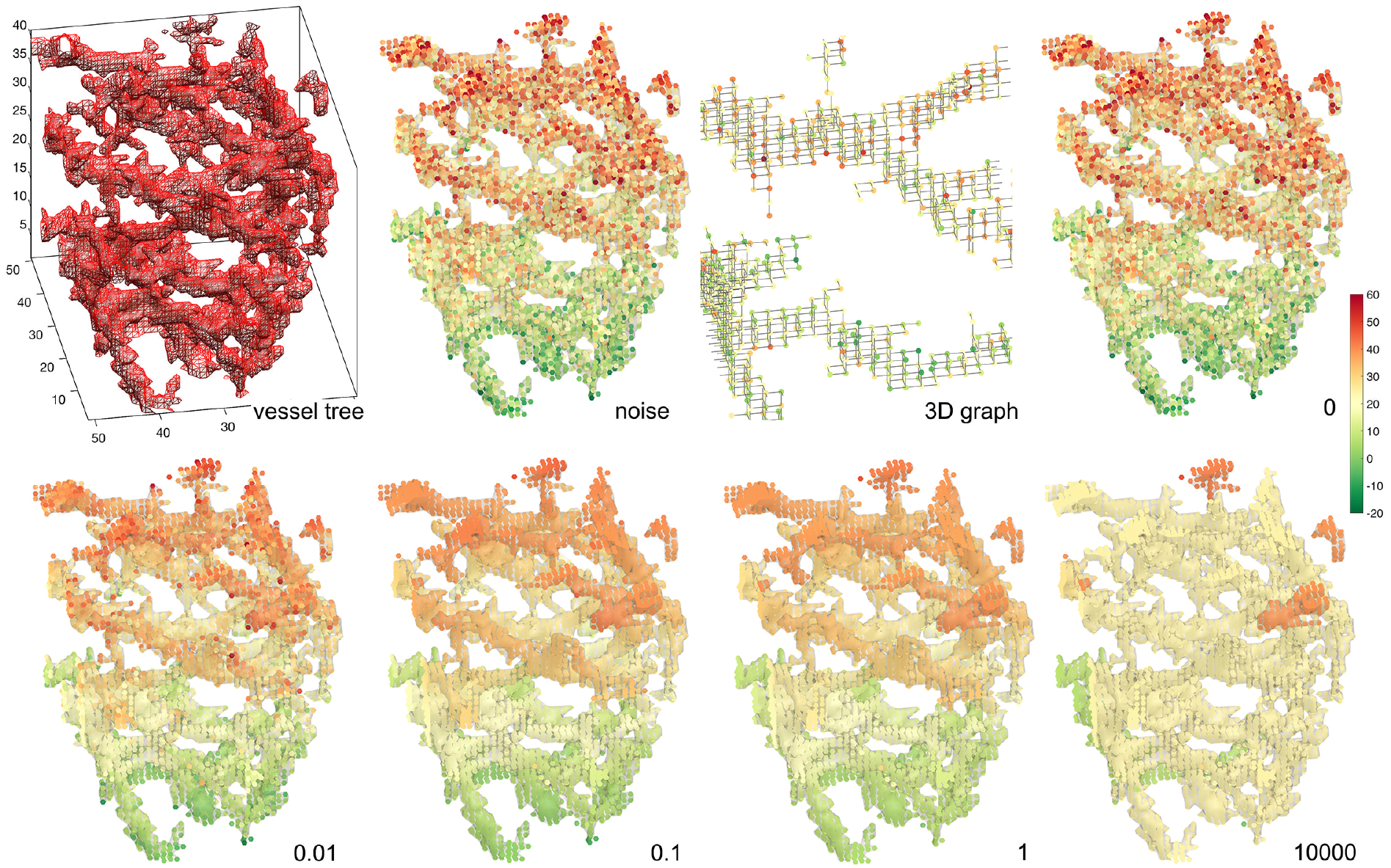}
\caption{From top left to right: 3D lung vessel tree. Gaussian noise is added to one of the coordinates. 3D graph constructed using 6-connected neighbors.  } 
\label{fig:lung-tree}
\end{figure*}

Define the $l$-norm of a vector $f=(f_1, \cdots, f_p)^\mathsf{T}$ as 
$$\parallel f \parallel_{l} =
            \Big(  \sum_{j=1}^p  \big| f_j \big|^{l}  \Big)^{1/l}. 
$$
The matrix $\infty$-norm is defined as
$$
\parallel f \parallel_{\infty} =
            \max_{1 \leq j \leq p}  \big| f_j \big|. 
$$

\noindent{\bf Theorem 1.} {\em Heat kernel smoothing is a contraction mapping with respect to the $l$-th norm, i.e.,
\bq \| K_{\sigma} *f \|_l^l \leq \|  f \|_l^l. \label{eq:contraction} \eq
}
{\em Proof.} Let kernel matrix $K_{\sigma} = (k_{ij})$. Then we have inequality
$$\| K_{\sigma} *f \|_l^l  = \sum_{i=1}^p  \sum_{j=1}^p |k_{ij}f_j |^l 
\leq \sum_{j=1}^p |f_j|^l.
$$
We used Jensen's inequality and doubly-stochastic property of heat kernel.

\section{Statistical properties}
Often observed noisy data $f$ on graphs is smoothed with heat kernel
$K_{\sigma}$ to increase the signal-to-noise ratio (SNR) and increases the statistical sensitivity
\cite{chung.2015.MIA}. We are interested in knowing how heat kernel smoothing will have effects on the statistical properties of smoothed data.

Consider the following addictive noise model:
\bqn f =    \mu + e, \label{eq:field} \eqn
where $\mu$ is unknown signal and $\epsilon$ is  zero mean noise.  Let $e= (e_1, \cdots, e_p)^\mathsf{T}$. Denote $\mathbb{E}$ as expectation and $\mathbb{V}$ as covariance. It is natural to assume that the variability of noises at different nodes $j$ is identical, i.e., 
\bqn \mathbb{E} e_1^2 = \mathbb{E} e_2^2 = \cdots = \mathbb{E} e_p^2. \label{eq:eeq} \eqn
Further, we assume that data at two nodes $i$ and $j$ to have less correlation when the distance between the nodes is large. 
So covariance matrix  $R_{e}  =  \mathbb{V} e = \mathbb{E} (e e^\mathsf{T}) = (r_{ij})$ can be given by
\bqn r_{ij} = \rho( d_{ij}) \label{eq:rij} \eqn
for some decreasing function $\rho$ and geodesic distance $d_{ij}$ between nodes $i$ and $j$. 
 Note
$r_{jj} = \rho(0)$ with the understanding that $d_{jj}=0$ for all $j$. The off diagonal entries of $R_e$ are smaller than the diagonal.

Noise $e$ can be further modeled as Gaussian white noise, i.e., Brownian motion or  the generalized derivatives of Wiener
process, whose covariance matrix elements are Dirac-delta. For the discrete counterpart, $r_{ij} = \delta_{ij}$, where $\delta_{ij}$ is Kroneker-delta with
$\delta_{ij}=1$ if $i=j$ and 0 otherwise.  Thus, 
$$R_e = \mathbb{E} (e e^\mathsf{T}) = I_p,$$
the identity matrix of size $p \times p$. Since $\delta_{jj} \geq \delta_{ij}$, Gaussian white noise is a special case of (\ref{eq:rij}).

Once heat kernel smoothing is applied to (\ref{eq:field}), we have
\bqn K_{\sigma}*f = K_{\sigma}*\mu + K_{\sigma}*e. \label{eq:Kfield}\eqn
We are interested in knowing how the statistical properties of data change from (\ref{eq:field}) to (\ref{eq:Kfield}). 
For $R_e = I_p$, the covariance matrix of smoothed noise is simply given as
$$R_{K_{\sigma}*e}  = K_{\sigma} \mathbb{E} (e e^\mathsf{T}) K_{\sigma} =  K_{\sigma}^2 =K_{2\sigma}.$$
We used the scale-space property of heat kernel. In general, the covariance matrix  of smoothed data $K_{\sigma}* e$ is
given by
\bq R_{K_{\sigma}* e} = K_{\sigma} \mathbb{E} (e e^\mathsf{T}) K_{\sigma}  = K_{\sigma} R_e K_{\sigma}.\eq

The variance of data will be often reduced after heat kernel smoothing in the following sense \cite{chung.2005.IPMI,chung.2005.NI}. This is formulated rigorously as follows. \\

\noindent{\bf Theorem 2.} 
{\em Heat kernel smoothing reduces variability, i.e., 
$$\mathbb{V} (K_{\sigma}* f)_j \leq \mathbb{V} f_j \label{eq:varreduction}$$
for all $j$. The subscript $_j$ indicates the $j$-th element of the vector.}

{\em Proof.} Note 
$$\mathbb{V} (K_{\sigma}* f)_j = \mathbb{V} (K_{\sigma}* e)_j = \mathbb{E} \Big( \sum_{i=1}^p k_{ij} e_i \Big)^2.$$
Since $(k_{ij})$ is doubly-stochastic, after applying Jensen's inequality, we obtain 
$$ \mathbb{E} \Big( \sum_{i=1}^p k_{ij} e_i \Big)^2 \leq \mathbb{E} \Big(  \sum_{i=1}^p k_{ij} e_i^2 \Big) = \mathbb{E} e_i^2.$$
For the last equality, we used the equality of noise variability (\ref{eq:eeq}). Since $\mathbb{E} f_j = \mathbb{E} e_i^2$, we proved the statement. 

\section{Application}

We applied heat kernel smoothing to the computed tomography (CT) of human lung obtained from DIR-lab dataset (\url{https://www.dir-lab.com}) \cite{castillo.2009, wu.2013}. Super-resolution procedure was used to enhance the resolution from  $1 \times 1 \times 2.5 \mbox{ mm}^3$ to $1 \times 1 \times 1 \mbox{ mm}^3$ resolution for CT images. The small part ($50 \times 50 \times 40$ voxels) of the CT image was used to illustrate the method and display the results better. The binary vessel segmentation was done using the multiscale Hessian filters at each voxel, as shown in the top left of Figure \ref{fig:lung-tree} \cite{frangi.1998,korfiatis.2011,shang.2011}. 

The binary segmentation was converted into a 3D graph by taking each voxel as a node and connecting neighboring voxels.  Using the 18-connected neighbor scheme, we connect two voxels only if they touch each other on their faces or edges.  If voxels are only touching at their corner vertices, they are not considered as connected. Although we used  the 18-connected neighbor scheme in this study, for visualization purpose only, Figure \ref{fig:lung-tree} displays the 6-connected neighbor scheme. This results in an adjacency matrix and the 3D graph Laplacian. Figure \ref{fig:lung-tree} example generates a 3D graph with  $p=6615$ nodes. The center of voxel is taken as the node coordinates. The large-scale eigenvector problem was subsequently solved using  an Implicitly Restarted Arnoldi Iteration method \cite{lehoucq.1996}. We used 6000 eigenvectors. Note we cannot have more eigenvectors than the number of nodes. Numbers in  Figure \ref{fig:lung-tree} are kernel bandwidths. At $\sigma=0$, heat kernel smoothing is equivalent to Fourier series expansion. Thus, we get almost identical results. As the bandwidth increases, smoothing converges to the mean value.
The regions that are different colors even with $\sigma=10000$ are regions that are disconnected. Thus, each disconnected regions are converging to their own different mean values.


\begin{figure}[t]
\centering
\includegraphics[width=0.8\linewidth]{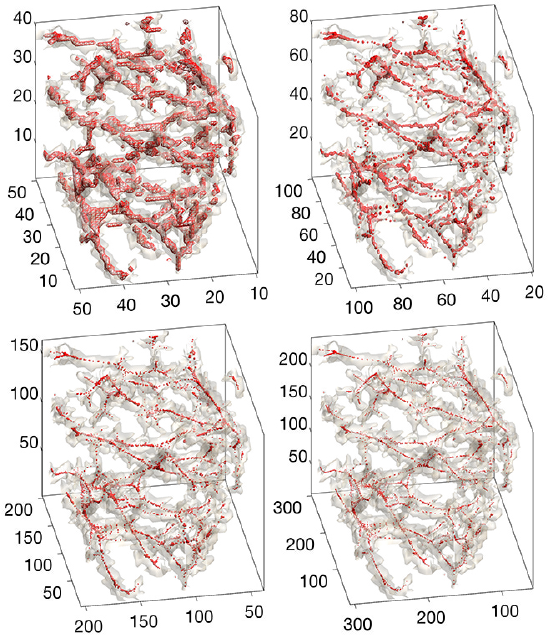}
\caption{The skeleton representation of vessel trees. Using the heat kernel series expansion with bandwidth $\sigma=1$ and 6000 basis, we upsampled the binary segmentation at 2, 4, 6 times (clockwise from top right) larger than the original size (top left).} \label{fig:lung-skel}
\end{figure}

\section{Discussion} 
The proposed technique can be used to extract the skeleton representation of vessel trees. Here we show the proof of concept. We perform heat kernel smoothing on node coordinates with $\sigma=1$. Then rounded off the smoothed coordinates to the nearest integers. The rounded off coordinates were used to reconstruct the binary segmentation. This gives the thick trees in Figure \ref{fig:lung-skel} (top left). To obtain thinner trees, the smoothed coordinates were scaled by the factor of 2,4 and 6 times before rounding off. This had the effect of increasing the image size relative to the kernel bandwidth (Figure \ref{fig:lung-skel} clockwise from top right). We hope that our method offers a new way of making the skeleton representation of the vessel trees \cite{lindvere.2013,cheng.2014}.

\section*{Acknowledgement}
This work was supported by NIH research grants R01 EB022856.  We would like to thank Jim Ramsay of McGill University and Michelle Carey of University College Dublin for valuable discussion on solving partial differential equations on irregular domains. We would like to thank Ruth Sullivan, Michael Johnson and Michael Newton of University of Wisconsin-Madison for useful discussion on modeling vessel trees.

\bibliographystyle{IEEEbib}
\bibliography{reference.2017.10.21}
\end{document}